# Pancomputationalism: Theory or Metaphor?

Vincent C. Müller
Anatolia College/ACT
www.sophia.de

17ᵗʰ November, 2008

The theory that all processes in the universe are computational is attractive in its promise to provide an understandable theory of everything. I want to suggest here that this 'pancomputationalism' is not sufficiently clear on which problem it is trying to solve, and how. I propose two interpretations of pancomputationalism as a theory: I) the world is a computer and II) the world can be described as a computer. The first implies a thesis of supervenience of the physical over computation and is thus reduced *ad absurdum*. The second is underdetermined by the world, and thus equally unsuccessful as theory. Finally, I suggest that pancomputationalism as metaphor can be useful. – At the Paderborn workshop 2008, this paper was presented as a commentary to the relevant paper by Gordana Dodig-Crnkovic.

## 1. Prelude: Some Science Fiction on *The Ultimate Answer* and *The Ultimate Question*

"Many many millions of years ago a race of hyperintelligent pan-dimensional beings (whose physical manifestation in their own pan-dimensional universe is not dissimilar to our own) got so fed up with the constant bickering about the meaning of life which used to interrupt their favourite pastime of Brockian Ultra Cricket (a curious game which involved suddenly hitting people for no readily apparent reason and then running away) that they decided to sit down and solve their problems once and for all.

And to this end they built themselves a stupendous super computer …

[…]

'O Deep Thought computer,' Fook said, 'the task we have designed you to perform is this. We want you to tell us …' he paused, 'the Answer!'

'The Answer?' said Deep Thought. 'The Answer to what?'

'Life!' urged Fook. 'The Universe!' said Lunkwill.

'Everything!' they said in chorus.

Deep Thought paused for a moment's reflection.

'Tricky,' he said finally.

'But can you do it?'

Again, a significant pause.

'Yes,' said Deep Thought, 'I can do it.'

'There is an answer?' said Fook with breathless excitement. 'A simple answer?' added Lunkwill.

'Yes' said Deep Thought. 'Life, the Universe, and Everything. There is an answer.

But,' he added, 'I'll have to think about it.'

[… at this point the whole procedure is interrupted by two representatives of the 'Amalgamated Union of Philosophers, Sages, Luminaries and Other Thinking Persons' who demand to switch off the machine because it endangers their jobs. They demand 'rigidly defined areas of doubt and uncertainty!', and threaten: 'You'll have a national Philosopher's strike on your hands!' This is resolved by Deep Thought who says it will take 7.5 million years to resolve the question and observes that, in the meantime 'So long as you can keep disagreeing with each other violently enough and maligning each other in the popular press, and so long as you have clever agents, you can keep yourself on the gravy train for life.' This convinces the philosophers and they leave. – 7.5 million years later, Deep Thought answers:]

'You're really not going to like it,' observed Deep Thought.

'Tell us!'

'All right,' said Deep Thought. 'The Answer to the Great Question …'

'Yes … !'

'Of Life, the Universe and Everything …' said Deep Thought.

'Yes … !'

'Is … ' said Deep Thought, and paused.

'Yes … !'

'Is … '

'Yes … !!! … ?'

'Forty-two,' said Deep Thought, with infinite majesty and calm.

[…]

'Forty-two!' yelled Loonquawl. 'Is that all you've got to show for seven and a half million years' work?'

'I checked it very thoroughly,' said the computer, 'and that quite definitely is the answer. I think the problem, to be quite honest with you, is that you've never actually known what the question is.'"

[Deep Thought then informs the researchers that it would design a second and greater computer, to tell them what the question is. That computer was built and called *The Earth*; many beings lived on it. Eventually, it was destroyed five minutes before the program was completed.]

(Adams 1979: 111-21, ch. 25-28)

… in a similar vein:

[The new Ultronic computer, with its $10^{6}$ logical units, is ready to answer any question. A little boy, Adam, asks "What does it feel like to be a computer?". The engineer says:]

"'Ultronic does not see what … it can't even understand what you are getting at!' The ripples of laughter about the room burst into a roar.

Adam felt acutely embarrassed. Whatever they should have done, they should not have laughed."

(Penrose 1989: 2, 583)

## 2. What is the Question? A Starting Point

Pancomputationalists say that the first story is literally true, in fact only part of the truth: not only is the Earth a computer, so is everything else; the universe is a computer. (I believe the term pancomputationalism was introduced in (Floridi 2004: 566).) In the following, I will try to learn a lesson from this strange story: Rather than investigate the truth of this answer, I will try to understand what it might mean – which will force me to speculate about what the question really is (a much harder problem than finding the answer, as we just learned).

A specification of what it means to say "everything is a computer" is particularly urgent because this view is in acute danger to be devoid of any meaning. This is not just the usual analytic philosopher's question "What do you mean?" – a very good question -, but a particular danger for any theory that says "everything is x": If everything really is x, then the defender of the theory cannot point to some samples that are x and then to other samples that are non-x to explain the theory. So, the defender has to explain under what conditions, counterfactually, something would not be x. If this is not done, the good old Karl Popper test of being in principle falsifiable is not passed and the theory is in acute danger to be devoid of any meaning. My impression is that the need for this explanation has been overlooked, in the enthusiasm about the explanatory power of the new all-encompassing theory.

The main question about pancomputationalism is thus what they might mean by saying that the universe is a computer. This is clearly not the only question, however, in particular I would expect there to be empirical questions to determine the truth of the theory.

It seems that there are two quite distinct traditions in pancomputationalism, namely a realist and an anti-realist one. Furthermore, the realist tradition involves theorists that use a very wide notion of computing, and others that understand the notion of computing in the traditional sense of *digital* computing, thus arguing for the stronger thesis that the world is ultimately digital.

### 2.1. Two Theories and a Metaphor

Imagine a ball bouncing up and down, finally coming to a rest. The pancomputationalist remarks: "This is all computation!" This remark has (at least) three interpretations:

A)  At any given time, the future states of the ball can be usefully described as the computational result from its present state (given all relevant factors).

B)  At any given time, the future states of the ball can be explained as the computational result from its present state (given all relevant factors).

C)  At any given time, the future states of the ball are computed from its present state (given all relevant factors). Bouncing is nothing but computation.

What A) says is just that under some meaning of 'computation' it may be useful to describe the process as computational – I call this "Metaphor" below.

What B) says is a deterministic physics is true and can be expressed mathematically – I call this "Theory II" below.

What C) says is that the bouncing of the ball can be reduced to computing. – I call this stronger view "Theory I" below.

### 2.2. Anti-Realist Pancomputationalism

A venerable tradition in the philosophy of computing answers the question 'which entities in the world compute?' with the remark 'It depends on how you describe it'. This view is sometimes called pancomputationalism because it says that anything in the world can be described as a computer, if we so please. Ver-sions of this tradition are represented, for example, by David Chalmers (Chalmers 1993; 1994; 1996) John Searle (Searle 1980; 1990; 1992: 207f) and Oron Shagrir (Shagrir 2006). As far as I can tell, this theory does not make any claims on the ultimate computational nature of the world or of the computational theory of the world; in fact it is often used to argue against a theory that the mind is computational in any substantial sense. In the following, I will deal with a stronger theory, what I call 'realist' pancomputationalism. (Perhaps my arguments are relevant for this weaker theory also, but I do not investigate this here.)

I happen to think that anti-realism is not the right stance in respect to the question which systems in the world are computers because I think that the underlying digital states can be individuated without invoking any observer or person with intentions, but this question is beside the current investigation (for the beginnings of an argument, see Müller 2008).

Note that "Theory II" is not a version of anti-realist pancomputationalism because the latter does not make a claim about the explanatory power of a description as computer.

### 2.3. Realist Pancomputationalism

There is an increasingly common view that the notion of computation can and should be used to describe many if not all physical processes; in fact that the physical world is at bottom computational. Gordana Dodig-Crnkovic says that, quite simply, "every natural process is computation in a computing universe" (Dodig-Crnkovic 2007: 10). Gregory Chaitin concurs: "The entire universe … is constantly computing its future state from its current state, it's constantly computing its own time-evolution! And … actual computers like your PC just hitch a ride on this universal computation!" (Chaitin 2007: 13f) Accordingly, the traditional foundational notions of matter and energy are supposed to be replaced by computation.

A new version of this theory is info-computationalism, namely the view that the physical universe *can best be understood* as *computational* processes operating on *informational* structure. Classical matter/energy in this model is replaced by information, while the dynamics are identified as computational processes. On the face of it, this thesis is stronger than mere pancomputationalism, since it is the conjunction of pancomputationalism with a claim about the substance on which computing takes place. There are also tendencies to express info-computationalism as an epistemic thesis, which would indicate that info-computationalism cannot be classified under the realist positions, after all. (See the paper by Dodig-Crnkovic in this volume.)

Within this framework, everything is computationalist. What is going on at the basic levels of physics and conventionally conceptualized, described, calculated, simulated and predicted in physics can be expressed in info-computationalist terms. Our bodies are advanced computational machines, at various levels, in constant interaction with other 'environmental' computational processes. Human nerve cells interact with each other and form complicated networks, thus producing another level of computational processes. At a yet higher level, these processes can be said to be processing representations of the outside world and result in events that are conscious to the agent, e.g. in what is called 'thinking'.

The traditional picture of computer science in general and of *artificial intelligence* in particular has been  that the intelligence of humans lies in these higher level cognitive or 'intellectual' facilities and that the aim of engineering is to reproduce these cognitive structures in a different computational hardware, e.g. in 'artificial cognitive systems'. It has become increasingly obvious in recent decades, however, that this approach has agents standing on their heads, instead of their feet. A successful natural intelligent agent is an organism with an evolutionary history that stands in a multitude of computational relations to its environment, including other agents. Many of these processes will make for complex systems and involve multiple agents or swarm intelligence. Pancomputationalism applies to the entire body of the agent, at various levels - the cognitive level is only one of many. Reproducing the computational structure would, on this picture, necessarily reproduce the 'emergent' properties of the natural agent. The wide pancomputationalist view will allow to understand and engineer a much larger range of processes than commonly included in computer science and artificial intelligence.

Which precise formulation of the theory one adopts also depends on the notion of 'computation' that underlies these theories. Some theorists use a very broad notion, while some rely on a notion of digital computing that is identical to Turing's (or at least very closely related to it). If digital computing is taken as a base, this leads to the idea of a "digital physics", based on an essentially discrete universe. For example: "everything is made out of 0/1 bits, everything is digital software, and God is a computer programmer, not a mathematician!" (Chaitin 2007: 3) This view goes back to John Wheeler's slogan "It from Bit" and is developed in computer science by Edward Fredkin (Fredkin 2003; 2007) and Stephen Wolfram (Wolfram 2002).

Pancomputationalism has repeatedly caught the public imagination, e.g. in *Wired* magazine (Kelly 2002) or at the recent very prominent Midwest KNS conference "What is computation? (How) does Nature compute?" (Bloomington, Indiana, 31.10.-2.11.08).

## 2.4. Computation

One of the many issues for pancomputationalism is to explain in which sense of computing the universe is a computer. We can, however, sidestep this issue for our argument. What we need is only the assumption that the pancomputationalist definition of computing, whatever it might be, will *include* classical Turing-machine computing. This kind of computing is a formal procedure that proceeds step-by-step and comes to a halt after a finite number of steps. The halting state is considered the 'output', the procedure that is followed is an algorithm. Such procedures that reach a halting state after a finite number of formal steps are also often called 'effective procedures'.

Having said that, it remains clearly a desideratum for the pancomputationalist to clarify what she means by computing – particularly for the realist version of the theory, since the anti-realist version is motivated by a particular view on computing already.

A further desideratum is to explain in what sense some physical systems, like the machine on which I am writing this, are clearly computers and some processes inside them are clearly computational processes – while other things (the apple next to it) are not, and yet others may be, e.g. the human nervous system. Pancomputationalism as theory would not make these disputes trivial (cf. Miłkowski 2007), but it would have to explain how some systems can be computers on the basis of computers.

## 3.   Pancomputationalism as a Theory I: The Universe Is a Computer

### 3.1. Two Forms of Ontological Reductionism

#### 3.1.1.   Strong (Type) Reduction

A version of reduction that is not uncommon in the natural sciences is the discovery that one known property (e.g. temperature) is identical to another, basic, property (e.g. kinetic energy) – but that talk in terms of the basic property has advantages in terms of explanatory power. In this sense:

Property A is reducible to basic property B if and only if:

> Necessarily, if two objects are identical with B, they are identical with A, and inversely.

If this is the case, any (extensional) talk in terms of properties A can be replaced by talk in terms of properties B *salva veritate*. We might want to say, for example that "Temperature is reducible to kinetic energy", in this sense. An "Identity theorist" in the philosophy of mind would say that "having a mental state of type A is reducible to having a brain state of type B" in this sense.

As the discussion of identity theory shows, this kind of reductionism goes beyond showing that property B is somehow 'basic' in claiming that there can be only *one way* in which the basic property can produce the property A. So, in our case, only one type of computation can produce one type of physical property. This seems implausible and as far as I can tell, it is not defended by pancomputationalists. In any case, it will be sufficient to investigate a thesis that is implied by this strong reductionist thesis.

### 3.1.2.  Global supervenience

One promising explanation of the pancomputationalist reduction thesis is that the physical properties are based on computational properties, without requiring that a particular computational property will always be basic for the same physical property. After all, the hope is that a few computational properties will explain a lot of physics. This kind of relation can be specified with a notion that has become increasingly popular in the philosophy of mind: the notion of supervenience. Here it is supposed to capture the intuition that the mental is based on the physical, in the sense that two physically identical objects must share the same mental properties, but without requiring that there is only one physical way to bring about a particular mental property (this will allow for the same mental properties in beings that are physically very different). Since physics is concerned with law-like explanations, it seems apt to formulate our thesis as a general statement about *types* of properties, rather than as about particular *tokens*. Here is a classical formulation of supervenience, taken from Jaegwon Kim:

> For families of properties A and B: "A *strongly supervenes* on B just in case, necessarily, for each x and each property F in A, if x has F, then there is a property in B such that x has G, and *necessarily* if any y has G, it has F." (Kim 1984: 165) – *weak* supervenience is defined just the same, but without the second 'necessarily'.

*Global supervenience* for two sets of properties M and P, where M properties supervene on the more basic P properties means that: "Any two worlds that are indiscernible with respect to their P properties are also indiscernible with respect to their M properties." (Shagrir 1999: 692)

This means that if two objects are identical with basic property P, they must be identical with M, *but not inversely*. In other words, there can be no difference in M without a difference in P. Or else: M is *multiply realizable* with different Ps. Classical examples for this kind of supervenience are: "Mental properties supervene on physical ones", "The look of a picture supervenes on its physical properties" (but the value does not, a physical duplicate of a painting by Rembrandt would look the same but have the value of a reproduction, not of a Rembrandt).

### 3.2. Reductive Info-Computationalism Produces Monsters

Let us now investigate the thesis that the physical processes are computational processes as saying that: "Physical processes supervene on computation". This means that if two physical processes $P_i$ and $P_j$ perform the same computation $C_l$, they are the same physical process.

This, however, seems to have absurd consequences. For one thing it would mean that reproducing a computational process is to reproduce the physical process: reproducing the hurricane in the computational model produces a hurricane!

Second, for all we know about computation, it is not true that there is no difference in M (derived) without a difference in P (basic). Computing is multiply realizable: $P_i$ and $P_j$ can be two *different* physical processes but both compute *the same* $C_l$. This contradicts supervenience of P on C – and it indicates that pancomputationalists may have supervenience upside down: if anything, computation supervenes on physics (but see below).

I think it is useful to put this point in terms of "levels of description" for a computer (cf. Floridi 2008; Müller 2008). A computer can be described on at least three levels:

- physical
- syntactic
- semantic

What we call computation takes place on the level of syntax, it is a purely formal procedure – taking place in a physical mechanism, and perhaps having meaningful symbols. Put in terms of these levels, what Searle stressed is that syntax does not determine semantics (semantics does not supervene on syntax) (Searle 1980). What I said above is that syntax does not determine physics (physics does not supervene on syntax).

David Deutsch, who uses the slogan "The world is made of qubits." as a version of Wheeler's "It from Bit" puts this point of multiple realizability thus: "Universality means that computations, and the laws of computation, are independent of the underlying hardware. And therefore, the quantum theory of computation cannot explain hardware. It cannot, by itself, explain why some things are technologically possible and others are not." (Deutsch 2004: 100, 01) Computation is not constrained enough to explain physical reality, what he calls 'the hardware'.

Incidentally, it is not true either that physics determines syntax, in the sense that computation would supervene on physics: $C_1$ and $C_2$ can be two different processes but both be computed by the same $P_1$. This can be a matter of function or interpretation (this insight got anti-realist pancomputationalism off the ground).

## 4. Pancomputationalism as a Theory II: A Complete Theory of the Universe Can Be Formulated in Computational Terms

Given the failure of taking pancomputationalism as a theory of the universe, it should be said that some formulations suggest it is more of a theory of an *explanation* of the universe – perhaps this is what the theory really says, rather than just being a consequence of the reductive theory? I will take a quick look at this possibility, though it is somewhat speculative.

Presumably what this theory might say is that the formal process of computing is sufficient for explanatory purposes, so it must claim that: "Any process is formally describable".

This would be good news for computer science, since it means that, in principle, anything can be programmed *perfectly*. Unlike in reductivist pancomputationalism this programming of some natural process is only a simulation – but there are no practical limits to what can be achieved in this way.

Whether this view is true is surely a deep question for the philosophy of mathematics, so allow me just to indicate why this position faces considerable obstacles.

Whatever precisely a formal description *is*, there are normally several possible formal descriptions for any given object or process. It is not clear that a particular one of these must be the right one (an impressive discussion is Putnam 1980). In particular, a formal description would have to specify 'what is the point' of the description, but normal physical processes do not have functions by themselves (with the exception of intentional processes and the possible exception of evolu-tionary processes), so the 'point' of the formalization must be provided by the observer.

For a computational description, this problem is probably even more serious, since even under a specific formal description of a process there are still several computations that it performs. If the description is "000110111001" and we are told in addition that the first 8 bits are 4 pairs of input and the last 4 are the output, then we might surmise that the process is addition. But this is underdetermined by the formal structure.

## 5. Pancomputationalism as Metaphor

What remains of pancomputationalism is its use as a metaphor, expressed in remarks like: "The universe can often usefully be described as computational", "The universe can often usefully be described as computing over information (as infosphere)", or "Some of our scientific knowledge can be described in computational terms". This metaphorical use will be a success if it is carefully distinct from more substantial philosophical theses. If this is ensured, it is very likely that many systems can usefully be described as computational, especially once a semiformal description of their relevant factors has been achieved. This metaphorical use is insightful and important – we should just not stretch it into a theory.